\documentclass[epj,english]{svjour}
\usepackage{graphics}
\begin{document}

\title{Multiphoton Ionization of Magnesium in a Ti-Sapphire laser field}

\author{
 L. A. A. Nikolopoulos\inst{1}, Gabriela Buic\u{a}-Zloh\inst{1,}\thanks
{permanent address: Institute for Space Sciences, P.O. Box MG-23, Ro 76900,
Bucharest-M\u{a}gurele, Romania},
and P. Lambropoulos\inst{1,}\inst{2} }

\institute{Institute of Electronic Structure and Laser, FORTH P.O.Box 1527,
 Heraklion-Crete 71110, Greece \and
 Physics Department, University of Crete, Heraklion-Crete, Greece}

\abstract{
In this paper we report the theoretical results obtained for partial
ionization yields and the above-threshold ionization (ATI) spectra
of Magnesium in a Ti:sapphire laser field (804 nm) in the range of
short pulse duration (20-120 fs). Ionization yield, with linearly
polarized light for a 120 fs laser pulse, is obtained as a function
of the peak intensity motivated by recent experimental data \cite{gillen:2001}.
For this, we have solved the time-dependent Schr\"{o}dinger equation
nonperturbatively
on a basis of discretized states obtained with two different methods;
one with the two-electron wavefunction relaxed at the boundaries,
giving a quadratic discretized basis and the other with the two-electron
wavefunction expanded in terms of Mg$^+$-orbitals plus one free
electron allowing the handling of multiple continua (open channels).
Results, obtained with the two methods, are compared and advantages
and disadvantages of the open-channel method are discussed.}

\authorrunning
\titlerunning
\maketitle
\date

\section{Introduction}
Experimental studies of above threshold ionization  have concentrated
mostly on the rare gases \cite{pet,scf,nan,buc} which can withstand
sufficiently large intensities of radiation of wavelength around 780 to 800
nm, which is the most versatile and convenient source
of short pulse duration.  As a result, extensive studies of ATI
photoelectron spectra, and of course high order harmonics,  in
the non-perturbative regime have and are being conducted
\cite{bur,corm,jian:1995a,jian:1996}.   Softer atoms,
with lower ionization potentials are not expected
to exhibit the extensive spectra of ATI, as they ionize at relatively lower
intensities.  Nevertheless, recent experimental studies
\cite{gillen:2001,druten:1994}
have addressed this issue in atomic Magnesium, which not only is a soft atom
but also has two valence electrons, which may or
may not add additional features, depending on intensity and pulse duration.
The results have been rather surprising.  The slopes of the ATI spectra as a
function of laser intensity are difficult to understand \cite{gillen:2001}.
 This motivated the theoretical study whose
main results are presented in this paper.  The study was undertaken because
we happened to have developed the theoretical tools necessary for a
realistic description of Magnesium as a correlated two-electron system.
Although under the conditions
of the experiment \cite{gillen:2001}, it does not appear that two-electron
excitations played
an important role, at least in the aspects we have
addressed, still we can say that the atomic structure entering the
calculation is realistic.  We have not calculated everything
that has been measured experimentally.  We present our results as a point of
calibration of what can be expected theoretically.
As discussed in the sections that follow, we find it difficult to understand
the observed slopes.  If it is because something is
missing in our calculation, at this point we are unable to say what it could
be.

\section{Two-electron continuum states of Magnesium}

The Hamiltonian of the Magnesium atom \( H_{a} \) can be written as:
\begin{equation}
\label{eq:h_2e}
H_{a}=\sum _{i=1}^{2}\left[ -\frac{1}{2}\nabla
_{i}^{2}+V_{eff}(\mathbf{r}_{i})\right] +V(\mathbf{r}_{1},\mathbf{r}_{2}),
\end{equation}
where \( V_{eff}(\mathbf{r}_{i}) \) represents the effective potential
for the \( i^{th} \) electron and the core, and \(
V(\mathbf{r}_{1},\mathbf{r}_{2}) \)
is a two-body interaction operator, that includes the static Coulomb
interaction $1/|{\bf r}_1 - {\bf r}_2|$ as well as a dielectronic
effective interaction \cite{chang:1993a,moccia:1996a}.

The characteristic feature of  Magnesium  is the existence
of an \( ns^{2} \) valence shell, outside a closed-shell core, the
excitation of which requires a much larger amount of energy compared
with the first and second ionization threshold. This allows us to
explore excitation and/or ionization processes of the valence electrons,
without considering the closed-shell core-excitation, for a certain
range of photon energies. The construction of the Magnesium bound and
final states is achieved through a standard Hartree-Fock calculation and
the inclusion of a core-po\-la\-ri\-zation potential which represents the influence
of the core on the two valence electrons in a way very similar to that described
in \cite{chang:1993a,nikolopoulos:2002a}.

At a first stage, we perform a Hartree-Fock calculation for the closed-shell
core of Magnesium (Mg$^{++}$), deriving thus the effective Hartree-Fock
potentials 'seen' by an outer electron. At a second stage, we solve
the Schr\"{o}dinger equation for  Mg$^{+}$. In this case,
the effective potential acting on the valence electron is given by \cite{moccia:1996a},
\begin{equation}
V_{eff}(r)=V_{l}^{HF}(r)+\frac{\alpha _{s}}{r^{4}}\left[
1-e^{-(\frac{r}{r_{l}})^{6}}\right],
\label{v_eff}
\end{equation}
where $\alpha_s$ is the static polarizability of the doubly-ionized Mg
and $r_l $ cut-off radii for the various partial waves $l = 0,1,2, ...$ .

Having produced the Mg$^+$  one-electron radial eigenstates $P_{nl}(r)$
for each partial wave $l = 0,1,2,...$, we solve the Schr{\"o}dinger
equation:\begin{equation}
\label{se_2e}
H_{a}\Psi ^{\Lambda }({\bf r}_{1},{\bf r}_{2}) = E \Psi ^{\Lambda }({\bf
r}_{1},{\bf r}_{2}),
\end{equation}
by expanding the two-electron eigenstates \( \Psi ^{\Lambda }({\bf r}_{1},{\bf
r}_{2}) \)
on the basis of two-electron antisymmetrized orbitals namely:
\begin{equation}
\label{eq:wf_2e_ci}
\Psi ^{\Lambda }({\bf r}_{1},{\bf r}_{2};E_{i}) =
\sum _{nll',n'}C_{nll'n'}(E_{i})\Phi _{nln'l'}^{\Lambda }({\bf r}_{1},{\bf
r}_{2})
\end{equation}

\begin{eqnarray}
\Phi ^{\Lambda }_{nln'l'}({\bf r}_{1,} {\bf r}_{2})& =& A_{12}
\frac{P_{nl}(r_{1})}{r_{1}}
\frac{R_{n'l'}(r_{2})}{r_{2}}\nonumber \\
&\times&
Y_{LM_{L}}(\hat{r}_{1},\hat{r}_{2};l,l')Y_{S,M_{S}}(s,s'),
\label{eq:wf_2e}
\end{eqnarray}
where \( A_{12} \) is the antisymmetrization operator which ensures
that the total wave function, is anti-symmetric with respect to interchange
of the space and spin coordinates of the two electrons. $\Lambda$ represents
the set of angular quantum numbers $(L S M_L M_S)$. In Eq.(\ref{eq:wf_2e})
the choice of the basis functions $R_{nl}(r)$ determines the method
used for the calculation of the correlated two-electron states.

In the case of fixed boundary conditions (FXBC), we force the wavefunction
to be zero at the boundaries by selecting the basis functions $R_{nl}(r)$ to
be the one-electron radial solutions of Mg$^{+}$, which by construction
vanish at the boundaries. This, transforms the Schr\"{o}dinger
Eq.(\ref{se_2e}) into a generalized eigenvalue matrix equation, by the
diagonalization of which we obtain the coefficients $C_{nll'n'}( E_i)$ for each
discrete eigenvalue $E_i$
\cite{chang:1993b,lambropoulos:1998,nikolopoulos:2002a}. This
choice of the basis functions $R_{nl}(r)$ leads thus to a discretized
continuum spectrum for the Magnesium atom with density of states determined
basically from the box radius $R$.

On the other hand, we can select the basis function $R_{nl}(r)$ to
be non-vanishing at the boundaries (FRBC) transforming the Schr\"{o}dinger
Eq.(\ref{se_2e}) into a system of algebraic equations for the
coefficients $C_{nll^{\prime}n^{\prime}}(E_i)$
\cite{lambropoulos:1998,nikolopoulos:2001a}. For
this, we use for both cases (FXBC and FRBC) the B-spline basis functions
\cite{deboor:1978} used extensively in the last 15 years in atomic structure
calculations. The main advantage of this method over the FXBC method is that
the density of the continuum spectrum is completely controllable (therefore
degeneracy into the continuum is guaranteed) and  the two-electron continuum
states have been expanded in terms of core-target states plus an ionized
electron, as in standard close-coupling scattering theory.

We have ascertained that the results obtained with the free boundary
method  compare well with the convergent results obtained with
the fixed boundary method.

\section{Time-dependent Schr\"{o}dinger Equation }

Our objective is to solve the time-dependent Schr\"{o}dinger equation (TDSE) :
\begin{equation}
\frac{d}{dt}\Psi ({\bf r}_{1},{\bf r}_{2};t) =-i \left[ H_{a} + V(t) \right]
\Psi ({\bf r}_{1},{\bf r}_{2};t).
\label{eq:tdse}
\end{equation}
The time-dependent interaction $V(t)$ of the two-valence electron atom with an
external laser pulse in the dipole approximation and velocity gauge can be written as:
\begin{equation}
V(t) = - A(t)\hat \varepsilon \cdot ({\bf p}_{1} + {\bf p}_{2}),
\label{eq:pulse}
\end{equation}
 where the vector potential \( A(t) \) is given by,
\begin{equation} A(t)=A_{0}f(t)\cos (\omega t) ,
\end{equation}
\({\hat \varepsilon } \) is the unit polarization vector of the laser field,
and \( {\bf p}_{1,2} \) are the momenta of the two valence electrons.
$A_0 =E_0/\omega$ is the amplitude of the vector potential, and \(
{\textrm{E}}_{0} \)
the electric field strength. The pulse shape envelope is \( f(t)=\cos
^{2}\left( {\pi t}/{\tau }\right)  \),
where \( \tau  \) is approximately the full width at half maximum  (FWHM) for
the electric field.
The integration time for the cosine squared shape pulse, is taken
from \( -\tau /2 \) to \( \tau /2 \). The velocity form of the interaction
Hamiltonian is chosen, because it makes the calculation converge faster,
in terms of the  number of  angular momenta included \cite{cormier:1996a}. It should
be noted here that length and velocity dipole matrix elements agree very well. For the
length operator, for the reason that the core-polarization potential is angular momentum
dependent, we have checked that the use of the modified operator as given by \cite{mitroy:1988},
${\bf r}^\prime = {\bf r} -{\bf \hat{r}} (\alpha_s/r^2)[1-e^{-(r/r_l)^6}]$ has very small effect in the
dipole matrix elements compared with those calculated with the standard length operator.

The time-dependent wavefunction is now expanded on the basis of eigenfunctions
\( \Psi ^{\Lambda }({\bf r}_{1},{\bf r}_{2};E_{i}) \),
\begin{equation}
\label{eq:wf_2e_t}
\Psi ({\bf r}_{1},{\bf r}_{2};t) =
\sum _{E_{i},L}U_{L}(E_{i},t)\Psi ^{\Lambda }({\bf r}_{1},{\bf r}_{2};E_{i}),
\end{equation}
 transforming the time-dependent Schr\"{o}dinger equation into a
set of first-order differential equations for the time-dependent coefficients
$U_L(E_i,t)$, namely,
\begin{equation}
\label{ode_u}
i \frac{d}{dt}U_{L}(E_{i},t)=\sum _{E_{i'}L'}[E_{i,L}\delta _{ii'}\delta _{LL'}
                                -V_{E_{i}L,E_{i'}L'}(t)]U_{L'}(E_{i'},t)
\end{equation}
subject to the initial conditions \( |U_{L}(E_{i},t=0)|^{2}=\delta
_{E_{i}E_{1}}\delta _{L0} \). The quantities \( V_{E_{i}L,E_{i'}L'}(t) \)
represent the matrix elements of \( V(t) \) calculated between the states characterized
by the quantum numbers \( E_{i}L \) and \( E_{i'}L' \). Assuming
that the laser field is linearly polarized, and the Magnesium atom
initially in its ground state \( (L=0,l_{1}=l_{2}=0,M=0) \), we need
to consider only the \( M=0 \) singlet states (\( S=0 \)).

At the end of the pulse, we obtain the coefficients $U_L(E_i,t\rightarrow
\infty)$, from
which we can extract information about the ionization yields, photoelectron
energy spectrum as well as photoelectron angular
distributions \cite{lambropoulos:1998,nikolopoulos:1999}.

\section{Results and discussion}
Compared to the Helium atom, the Magnesium continua associated with
more than one ionization threshold lie much closer to the lowest one
(one open channel) which makes them much more accessible at intensities,
frequencies, and pulse durations currently available. That is why
excited ionic states have often  been observed in alkaline earth atoms.

\begin{figure}
\centerline{\resizebox*{10cm}{!}{\includegraphics{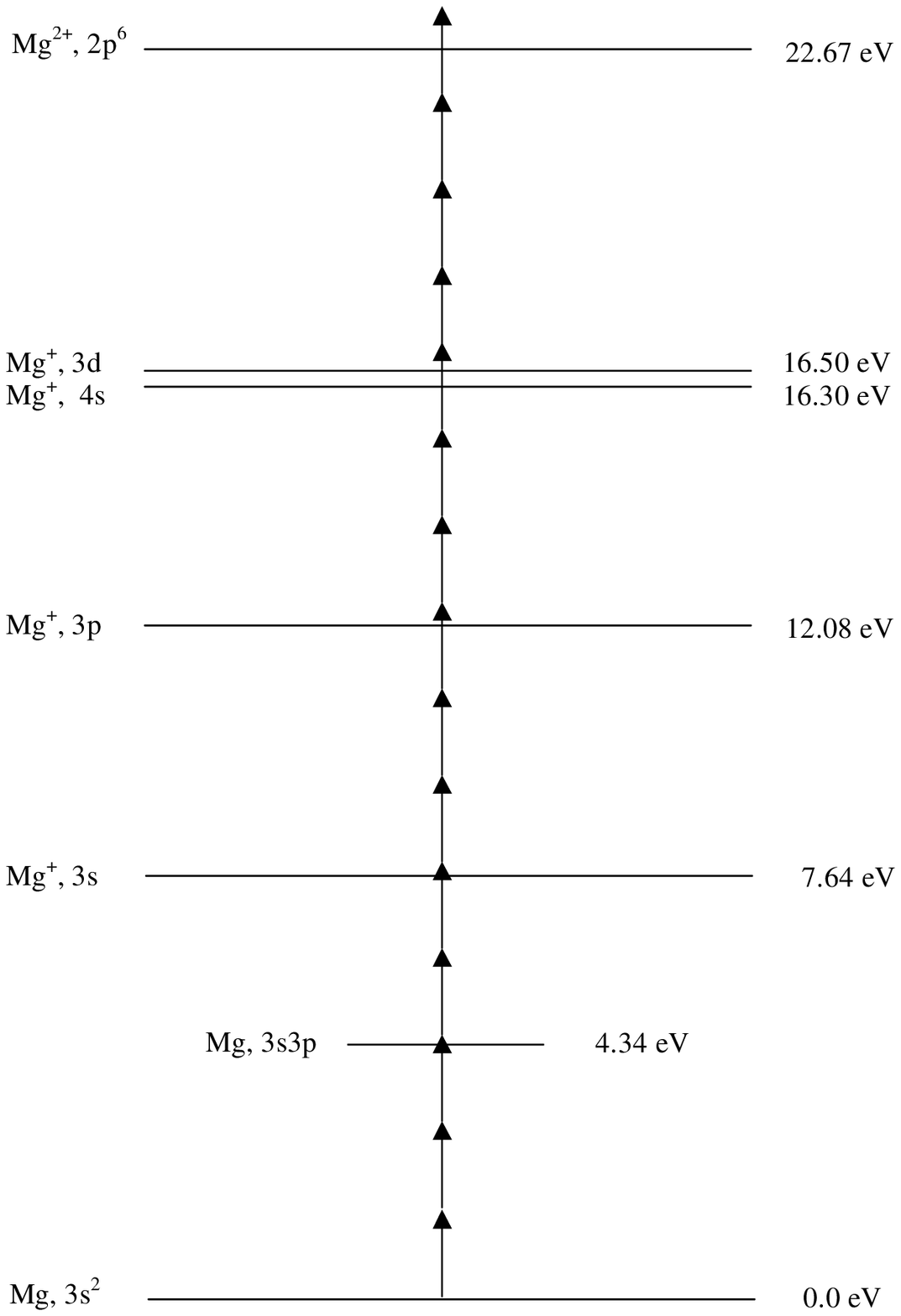}} }
\caption{\label{fig:fig1}}\vspace*{1cm}
\end{figure}

For laser intensities of about $I_0<10^{12}$ W/cm$^2$ of a Ti:sapphire
laser at 804 nm (1.54 eV), five photons are required to ionize the
Magnesium atom
\begin{equation}
{\rm Mg}(3s^2)+5\omega \rightarrow {\rm Mg}^+(3s)+e^-(0.06 \;{\rm eV}),
\end{equation}
and leave the Mg$^+$ in its ground state, as illustrated in  the energy-level
diagram  in Figure \ref{fig:fig1}. At higher intensities as the
 ponderomotive shift increases,
at least six photons are required to ionize the Magnesium atom and
the \( 3s3p \) state can be reached through a three-photon absorption.
If we consider, for example, 11-photon ionization in Magnesium, five
of the photons will bring the atom in the ground ionic state plus
one electron in the associated continuum, and the remaining ones will
be absorbed with the possibility to excite further the ionized electron
(above-threshold ionization) and/or to excite the core to different
ionic states. The calculation of the probability of this channeling
into the final ionization stages of the Magnesium, after the interaction
with the laser, is a difficult task. In the following paragraphs, we
discuss the results obtained with two different methods each
 having its own advantages and disadvantages.

Populating the first excited ionic state \( {\textrm{Mg}}^{+}(3p) \)
is possible by the absorption of eight photons. With the absorption
of an additional three photons, the ionic excited states \(
{\textrm{Mg}}^{+}(4s) \)
and \( {\textrm{Mg}}^{+}(3d) \) become also energetically accessible.
The position in energy of the ionic thresholds in Figure \ref{fig:fig1},
is for the bare atom without taking into account the shift of the energy
levels in the laser field. In the presence of an intense laser,
the ionization threshold of an atom is increased by the ponderomotive energy.
If the intensity of the laser is sufficiently
high, channel closing will occur so that \( N\hbar \omega <I_{p}+U_{p} \),
where \( I_{p}=-E_{a} \) is the ionization potential corresponding
to the energy of the ionization threshold \( E_{a} \), \( N \) is
the number of photons required to ionize the atom, and \( U_{p} \)
is the ponderomotive energy which is given in atomic units by \(
U_{p}={\textrm{E}}_{0}^{2}/4\omega ^{2} \).
In the case of photon energy \( \omega =1.54 \) eV the ponderomotive
shift is \( U_{p}\simeq  \) 0.6 eV at the intensity \( I \)=10\(^{13} \)
W/cm\(^{2} \).

A related phenomenon which occurs in strong laser fields is that large
AC-Stark shifts in atomic states can lead to intermediate resonances
in the ionization process \cite{agostini:1989}. At intensities around
\(2.4\times 10^{12} \) W/cm\(^{2} \), the excited state of the Magnesium atom
\( 3s3p \) is shifted into resonance.

\subsection{Fixed boundary condition method}

We have studied the intensity dependence of both ion yield and the
photoelectron spectrum of Magnesium in the Ti-Sapphire laser field
in the domain of short pulse duration for laser intensities up to
\( 10^{14} \) W/cm\( ^{2} \) with two-electron states of Magnesium
 calculated with the fixed boundary method. The results presented
are converged in terms of box size,  number of B-splines and angular
momenta.

\begin{figure}
\centerline{\resizebox*{8cm}{!}{\includegraphics{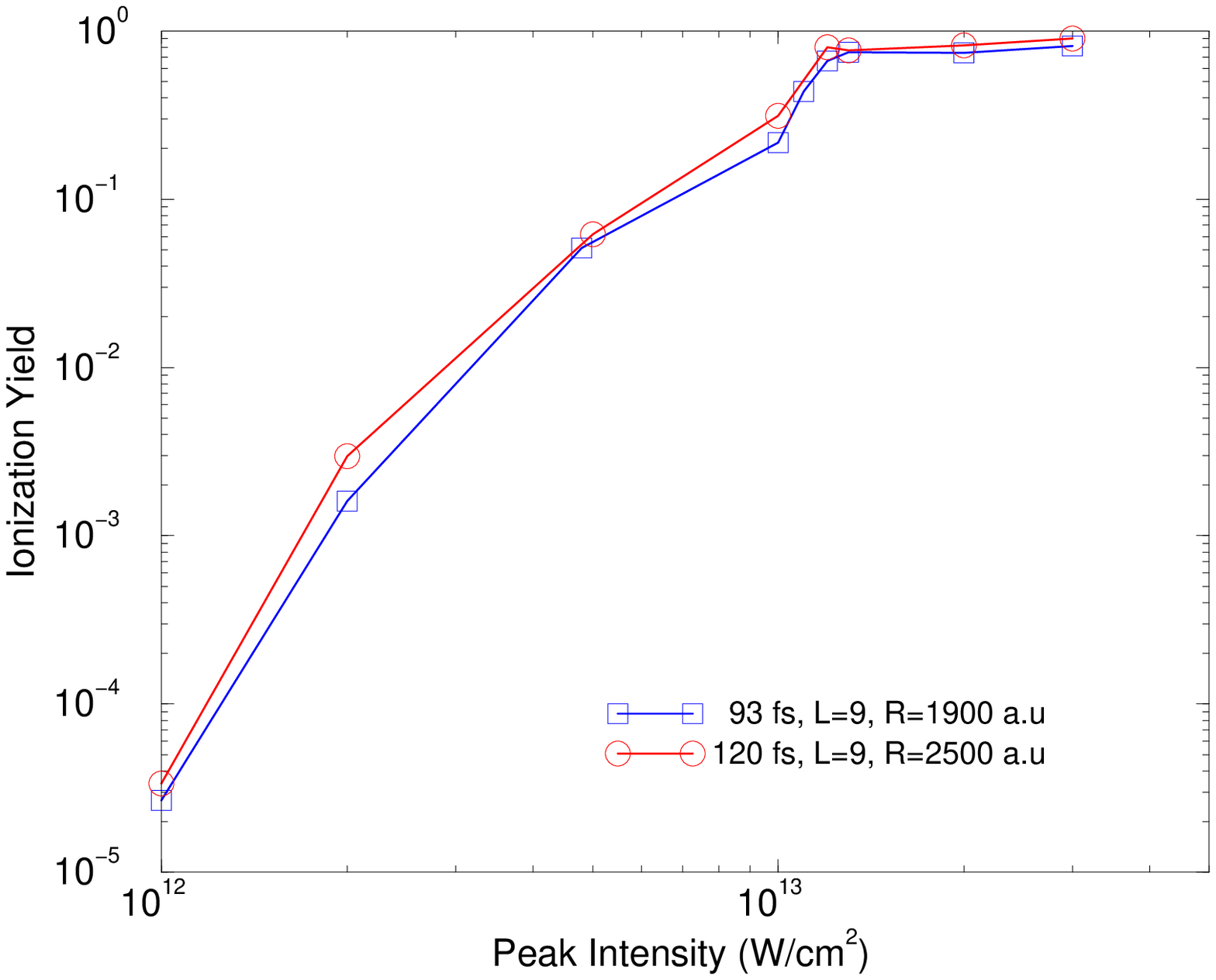}} }
\caption{\label{fig:fig2}}\vspace*{1cm}
\end{figure}

In Figure \ref{fig:fig2}, we show the yield of Magnesium
ions as a function of the laser intensity, for photon energies
 \( \omega =1.54 \)
eV, and laser pulse duration \( \tau =35 \)  optical cycles ($\sim 93$ fs)
and \( \tau =45 \)  optical cycles ($\sim 120$ fs).
We have employed  a basis of \( 992 \) B-splines in a box of \( 1900 \) a.u,
and a  total number of angular momenta up to L = 9.

According to perturbation theory, the \( N \)-photon generalized
cross section is proportional to the \( N^{th} \) power of the photon
flux. Therefore the slope of the ionization yield versus laser intensity
in a logarithmic scale represents the minimum number of photons necessary
to ionize the atom. Since the Magnesium states in the continuum are
shifted in the laser field with the ponderomotive potential, at least
six photons must be absorbed to ionize the Magnesium atom for laser intensities
higher than \( 10^{12} \) W/cm\(^{2} \).

For intensities in the range between $10^{12}$ W/cm$^{2}$ and
 $10^{13}$ W/cm$^2$, and for both the 93 and 120 fs pulses we found slopes
for the ionization yield curves around  6 which are basically in agreement
with the lowest-order perturbation theory (LOPT) predictions. The
ionization yield of Mg$^+$ increases linearly with the peak intensity
of the field until the saturation regime is reached, where a decrease
in the slope begins. Around $10^{13}$ W/cm$^2$, however, we have found a
sudden enhancement of the ionization yield which is attributed to AC-stark
shifting of the Rydberg states into resonance with the laser field. Though this
enhancement is apparent in our results, the slope does not compare to that
(of about $ 11$)  reported in the recent experiment by \cite{gillen:2001},
 in the same range of intensities.

\begin{figure}
\centerline{\resizebox*{8cm}{!}{\includegraphics{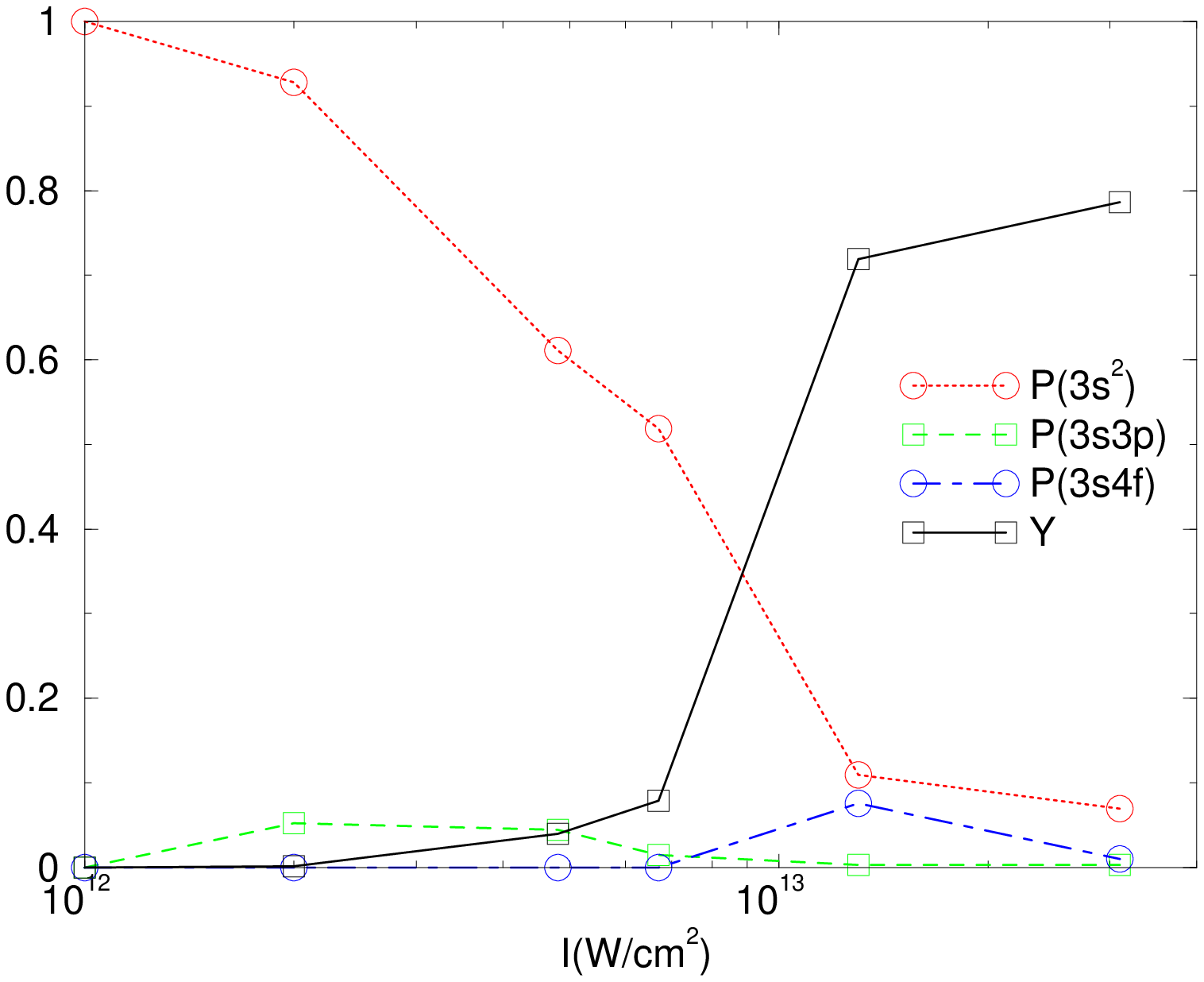}} }
\caption{\label{fig:fig3}} \vspace*{1cm}
\end{figure}

\begin{figure}
\centerline{\resizebox*{8cm}{!}{\includegraphics{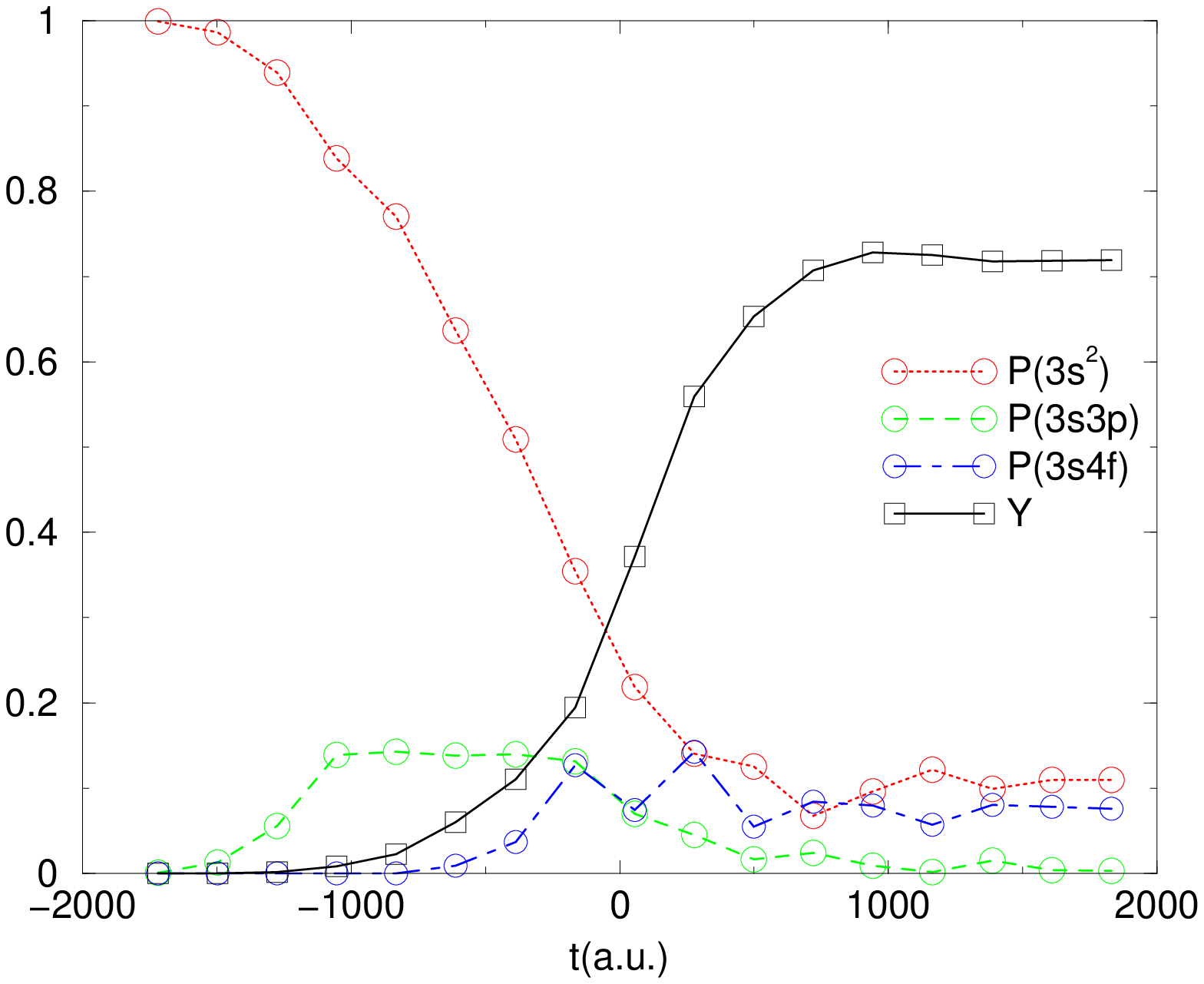}} }
\caption{\label{fig:fig3.1}} \vspace*{1cm}
\end{figure}

In Figure \ref{fig:fig3} we have plotted the ground state population, the ionization
yield as well as the population of states $3s3p$ and $3s4f$. In this figure one
can see that at peak intensity of about $1.3\times10^{13}$ W/cm$^2$ the ionization yield
makes an abrupt 'jump'. At the same intensity the population of the state
$3s4f$ is almost 10$\%$ of the ionization yield (almost equal with the remaining
population in the ground state) while the population of the $3s3p$ state becomes
negligible. In Figure \ref{fig:fig3.1} we have plotted for the peak intensity
$1.3\times10^{13}$ W/cm$^2$ the populations of the same states as in Figure \ref{fig:fig3}
and the ionization yield during as a function of the time. The $3s3p$ state is populated during
the rise of the pulse while at the end of the pulse the population of the $3s3p$ state has been
'transferred' to the $3s4f$ state, which becomes the dominant, in population, excited state.

%
% Given that, the calculation of an ATI spectrum is much more demanding than the
% calculation of the ionization yield at given laser intensity, in terms of
% convergence criteria, we plot the photoelectron energy spectrum, calculated
% at the end of the laser pulse(Figure \ref{fig:fig3}) when the duration of the
% pulse is 20 optical cycles ( $\sim$ 53 fs) and the photon wavelength  804 nm.
% The peak laser intensity is \( I_{0}= 2\times 10^{12} \) W/cm\( ^{2} \).
%

\subsection{Free boundary condition method}

We produce the initial ground state of Magnesium in a box of radius
\( R=40 \)~a.u. in which we have included only 20 configuration
series of the type \( nsms,npmp,ndmd,nfmf \) with \( n,m \) up to
5. The number of B-splines basis was 62 and the order 9. The parameters
of the core-polarization potential included in Eq.\ref{v_eff} were
 \( \alpha_s = 0.491 \)
for the static polarizability of the closed-shell core and 1.2, 1.335,
1.25, 1.3, 1.1 for the cut-off radii for the \( l=0,1,2,3,4, \) partial
waves, respectively. This resulted to a value for the singly ionized
Magnesium equal $ E$(Mg\( ^{+}(3s))= - 14.99 \) eV and a value for the
Magnesium ground state equal to \( -22.653 \) eV measured from the double
ionization threshold. The energy value for the intermediate $3s3p$ state was
$-18.32$~eV.

The basis used for the propagation of the TDSE differs from
the one used in the FXBC case in two aspects: firstly, we have produced
our basis in a much smaller box (of radius 40 a.u) than that of the
FXBC basis ($\sim 600 \div 2000$ a.u.) which results in the absence of high
Rydberg states in the Mg bound spectrum. Thus, while in the FXBC basis
a number of Rydberg states of about $\sim 30 \div 40$ might be obtained,
depending on the value of the box radius, in the FRBC basis a number
of 5 Rydberg states is obtained. Moreover, we have calculated a
total number of  angular momenta, up to five. The only reason
for this restriction of the FRBC basis is computational resources
limit. It is the price that one has to pay in going from
the FXBC method to the FRBC method. The expansion of the two-electron
states in a set of target - core states plus one free electron,
allows us to obtain  much more detailed information for the states of
the atom after its interaction with the laser, such as partial ionization
yields (populations of the various ionization stages of the ionized
system), partial photoelectron spectra (branching ratios) and moreover
angular photoelectron distributions. Thus it is expected, that quantitative
differences are to be found between the two approaches due to the limited
basis used in the FRBC approach. It is absolutely necessary, therefore,
that one should ensure that results  calculated with both methods are
in good agreement, since in an unlimited basis case they have to be the same.

\begin{figure}
\centerline{\resizebox*{8cm}{!}{\includegraphics{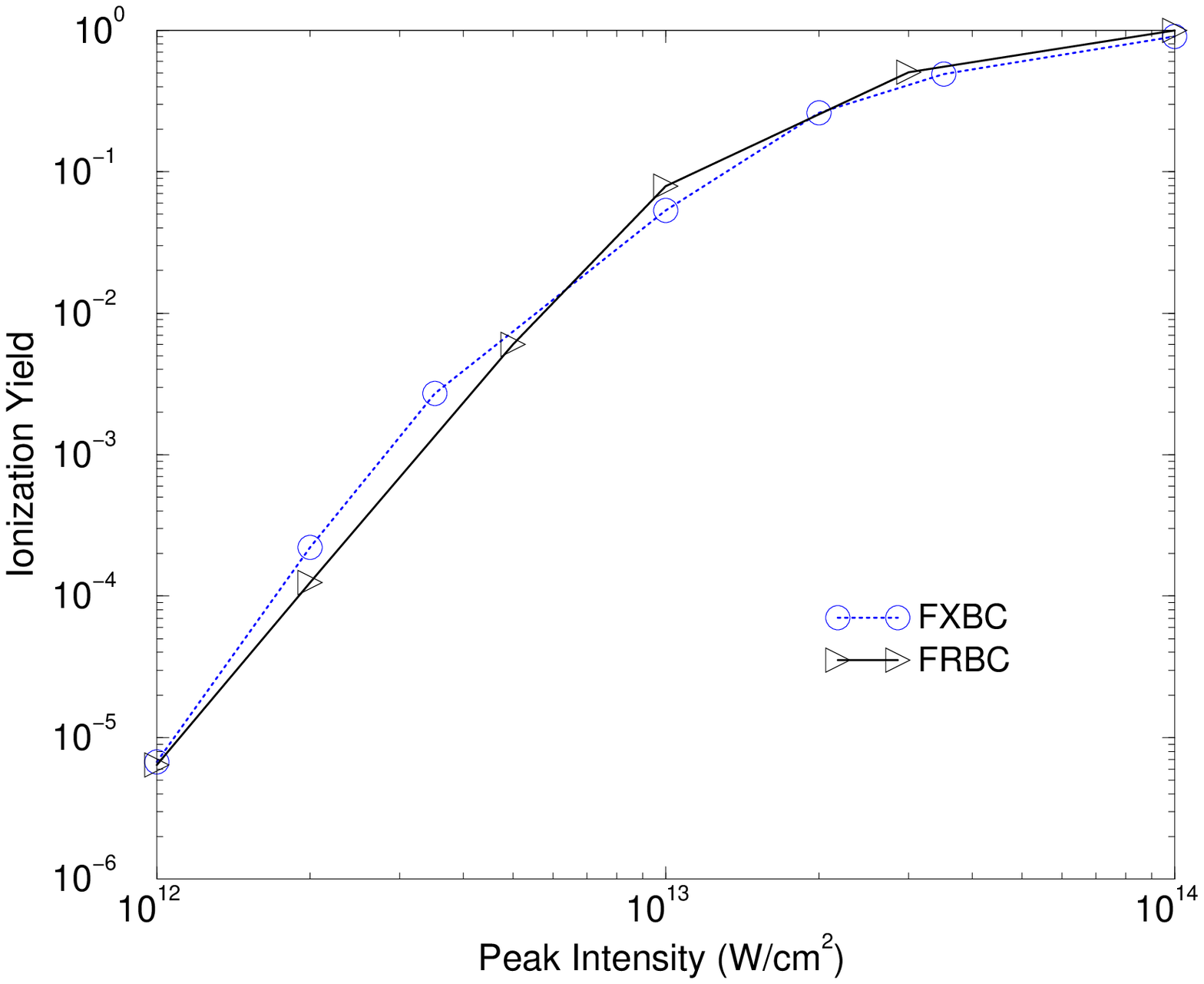}} }
\caption{\label{fig:fig4}} \vspace*{1cm}
\end{figure}

With this in mind, we have compared in Figure \ref{fig:fig4} the ionization yield
in Magnesium calculated with the free boundary method for a pulse
duration of about 29.5 fs (full curve) with the ionization yield calculated
with the fixed boundary method of about 27 fs (dotted curve). We have
employed a basis of \( 802 \) B-splines in a box of \( 900 \) a.u and
calculated two-electron states with total angular momenta up to L = 9, for the
results obtained with the fixed boundary method. For short pulse duration,
the agreement between these two methods is rather good. The corresponding
slope of the ionization yields is 4.46 and 4.25 for the FXBC and FRBC
method, respectively. Similar agreement is obtained for the photoelectron
energy spectrum for the pulses employed here, though not presented.

In Figure \ref{fig:fig5}, we have plotted the ion yield versus intensity
of the laser field, of Magnesium ions left in different states, for
photon energy \( \omega =1.54 \) eV and 93 fs pulse duration, calculated
with a cosine squared pulse. From the ionization yield curves, we
found that the Mg ion is preferentially left in the \( 3s \) ground
state in the intensity domain smaller than \( 2\times 10^{13} \)
W/cm\( ^{2} \). The slope of the corresponding curve (represented
by the dotted line) has the value 5.96 around the peak intensity \(
I_{0}=10^{12} \)W/cm\( ^{2} \). We should mention here the difference in
the slope of the ionization yield between the shorter pulse (fig. \ref{fig:fig4})
which gives a slope of about $\sim 4.4$. For the longer pulse the photon energy
spectrum of the field is much less broader than in the case of the shorter pulse.
Due to bandwidth of the shorter pulse, a three photon absorption from the ground state
populates strongly the 3s3p excited state which for that reason, plays the role
of an intermediate resonance thus lowering the order of the slope of the ionization yield,
as expected from energy conservation rules, according LOPT predictions.
Note that, ionization from the 3s3p state requires the absorption of three more 1.55 eV photons.
Similar role plays the ionization threshold at about $\sim 7.64$~eV, while the 5-photon
(1.55 eV) absorption brings the system in an energy very close to the ionization threshold.

In the presence of resonances, the ionization rate for multiphoton
processes measured as a function of intensity will deviate from the
power-law \( I^{N} \) dependence \cite{lambropoulos:1976}. Since
we are dealing with near resonant multiphoton ionization with the
\( 3s3p \) state
of Magnesium, the slopes of the ionization yields for the Mg\( ^{+}(3p) \),
Mg\( ^{+}(4s) \), and Mg\( ^{+}(3d) \) thresholds are smaller than
those predicted by  perturbation theory. In order to have a better
picture of the amount of ions left in different excited states, we
have calculated the branching ratios in photoelectron spectra corresponding
to different ionization channels and their dependence on laser intensity.
In Figure \ref{fig:fig6}, we have plotted the branching ratios of partial
ionization yields, namely the number of ions left in a particular
state of the Magnesium ion (ground or excited). The branching ratios have
been calculated by integrating over the energy the partial photoelectron
spectrum associated with that state. It is well-known that, within
 lowest-order  perturbation theory, the branching ratios are independent
of laser intensity, as we can observe at intensities less than \( 2\times
10^{12} \)
W/cm\( ^{2} \) for \( 3p \) and \( 4s \) ionization thresholds.
At moderate intensities, the number of ions left in Mg\( ^{+}(3p) \)
state, represented by the dotted curve, is greater than the ions left
in the Mg\( ^{+}(3d) \) state (dot-dashed curve) and Mg\( ^{+}(4s) \)
state (dashed curve). For higher intensities, in the saturation
domain, the amount of ions left in the Mg\( ^{+}(3p) \) and Mg\( ^{+}(3d) \)
excited states are of comparable value.

We also present in Figure \ref{fig:fig7}, the partial photoelectron
energy spectra corresponding to ions left in the ground state (full
curve) and in different excited states: \( 3p \) (dotted curve),
\( 3d \) (dashed curve), \( 4s \) (dot-dashed curve) calculated for
a peak intensity of the laser pulse \( 2\times 10^{12} \) W/cm\( ^{2} \).
The spectrum associated with the \( 3p \) threshold is hidden under
the dominant contribution associated with the \( 3s \) threshold. This
overlap can be attributed to the fact that the \( 3s \) and \( 3p \)
thresholds are about \( 4.44 \) eV apart, so that the first photoelectron
peak above the \( 3s \) threshold has only \( 0.18 \) eV energy
less than the first peak above the \( 3p \) threshold.

\section{Conclusion}
In conclusion, we have calculated ionization yields and ATI spectra
of Magnesium subject to a pulse of duration of 93 fs and 120 fs for
the Ti:sapphire laser (804 nm) for the range of intensities between 10$^{12}$
W/cm$^{2}$ and 5$\times$ 10$^{13}$ W/cm$^2$. For this,
we have employed a well established method, based on a configuration
interaction discretized two-electron states, developed over
the last few years \cite{jian:1995a,jian:1996,nikolopoulos:1999,tang:1990,tang:1991},
and in addition we have presented an alternative method to solve
nonpertubartively
the time-dependent Schr\"{o}dinger equation that handles ionization stages of
two-electrons atoms with multiple continua (many open channels).
Partial ionization yields, photoelectron spectra and photoelectron
angular distributions (though not presented in this work) may be obtained
with this open-channel method.

As already mentioned in the text, the slopes of the ion yield as a function of laser
intensity that we obtain are considerably lower than those observed by Gillen et al
\cite{gillen:2001}.
 Our calculations have not included averaging over the interaction volume.  Such
averaging in any case tends to lower rather than increase the slopes of the ion yield curves.
 The effect of resonance with intermediate atomic states is inherent in our calculations.
 But in any case, it is not expected to raise slopes.  Thus at this point, we have no
clue as to the possible reasons for the discrepancy.
 Hopefully, further experimental and possibly theoretical work will shed some light
on this puzzle.

\begin{acknowledgement}
 The work of G.B-Z. was supported by the European Research network
Program contract No HPRN-CT-1999-00129. One of the authors (G.B-Z.) is
indebted to Dr. Takashi Nakajima for useful discussions.
\end{acknowledgement}

\newpage

{\centering TABLE - FIGURE CAPTIONS }

FIG. 1. Energy-level diagram for Magnesium showing the levels relevant
for this study. The energies are referenced with respect to the neutral
Magnesium ground state.

FIG. 2. Ionization yield in Magnesium as a function of peak laser intensity
for photon energies \( \omega =1.54 \) eV. Pulse duration is  93 fs (35 optical cycles)
and 120 fs (45 optical cycles).

FIG. 3. Population of $3s^2, 3s3p, 3s4f$ states and ionization yield (Y), at the end of the pulse,
 as a function of the peak laser intensity, for the 93~fs pulse.

FIG. 4. Population of $3s^2, 3s3p, 3s4f$ states and ionization yield (Y) as a function of the interaction
time, for the 93~fs pulse, and peak intensity $1.3\times 10^{13}$W/cm$^2$.

FIG. 5. Comparison of ionization yields in Magnesium calculated with
FRBC method (full curve) and FXBC (dotted curve) method as a function
of the peak laser intensity for photon energy at 804 nm (1.54 eV).

FIG. 6. Ionization yield in Magnesium as a function of peak laser
intensity for photon energies \( \omega =1.54 \) eV. Pulse duration
is 93 fs (35 optical cycles).

\begin{figure}
\centerline{\resizebox*{8cm}{!}{\includegraphics{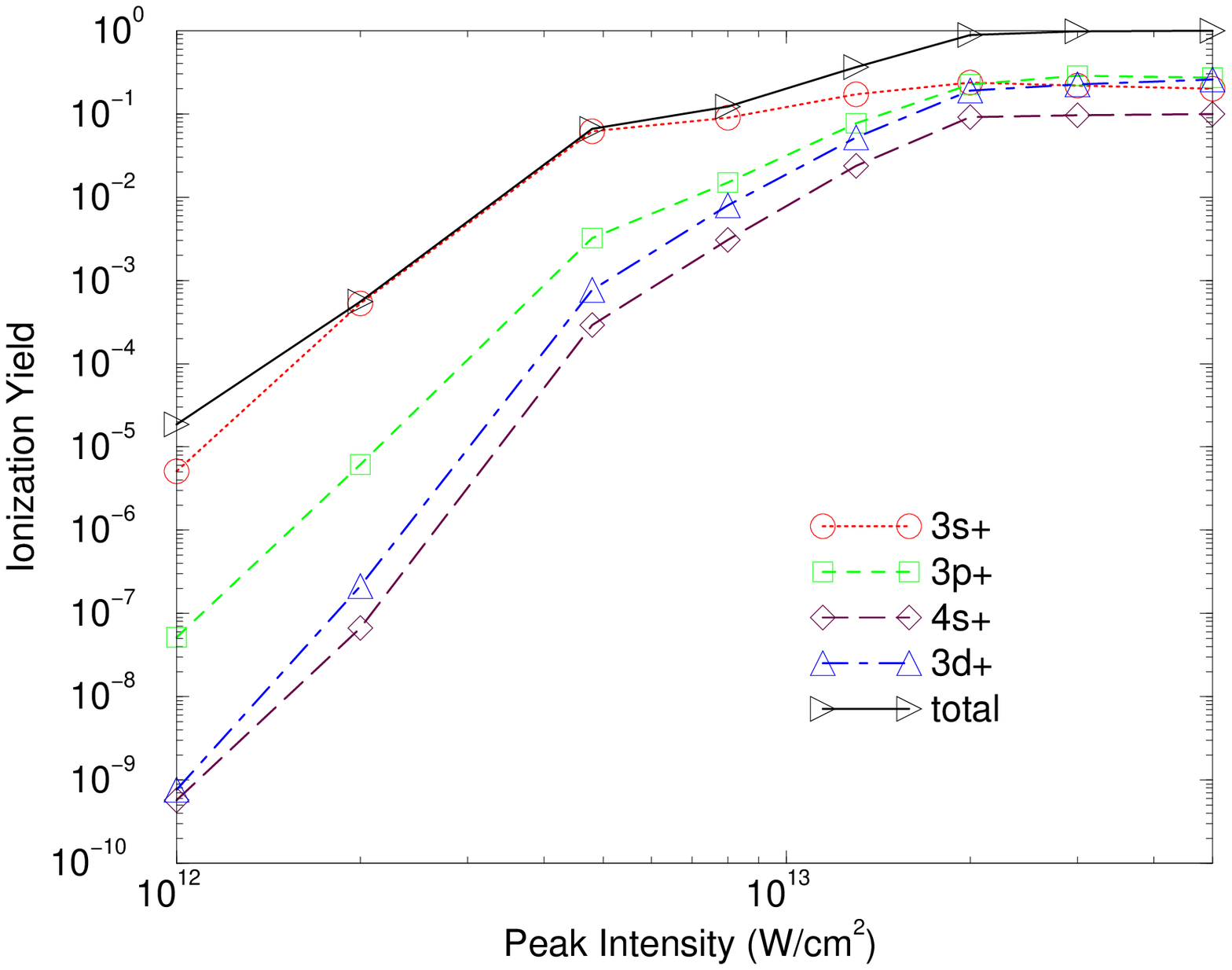}} }
\caption{\label{fig:fig5}}\vspace*{1cm}
\end{figure}

\begin{figure}
\centerline{\resizebox*{8cm}{!}{\includegraphics{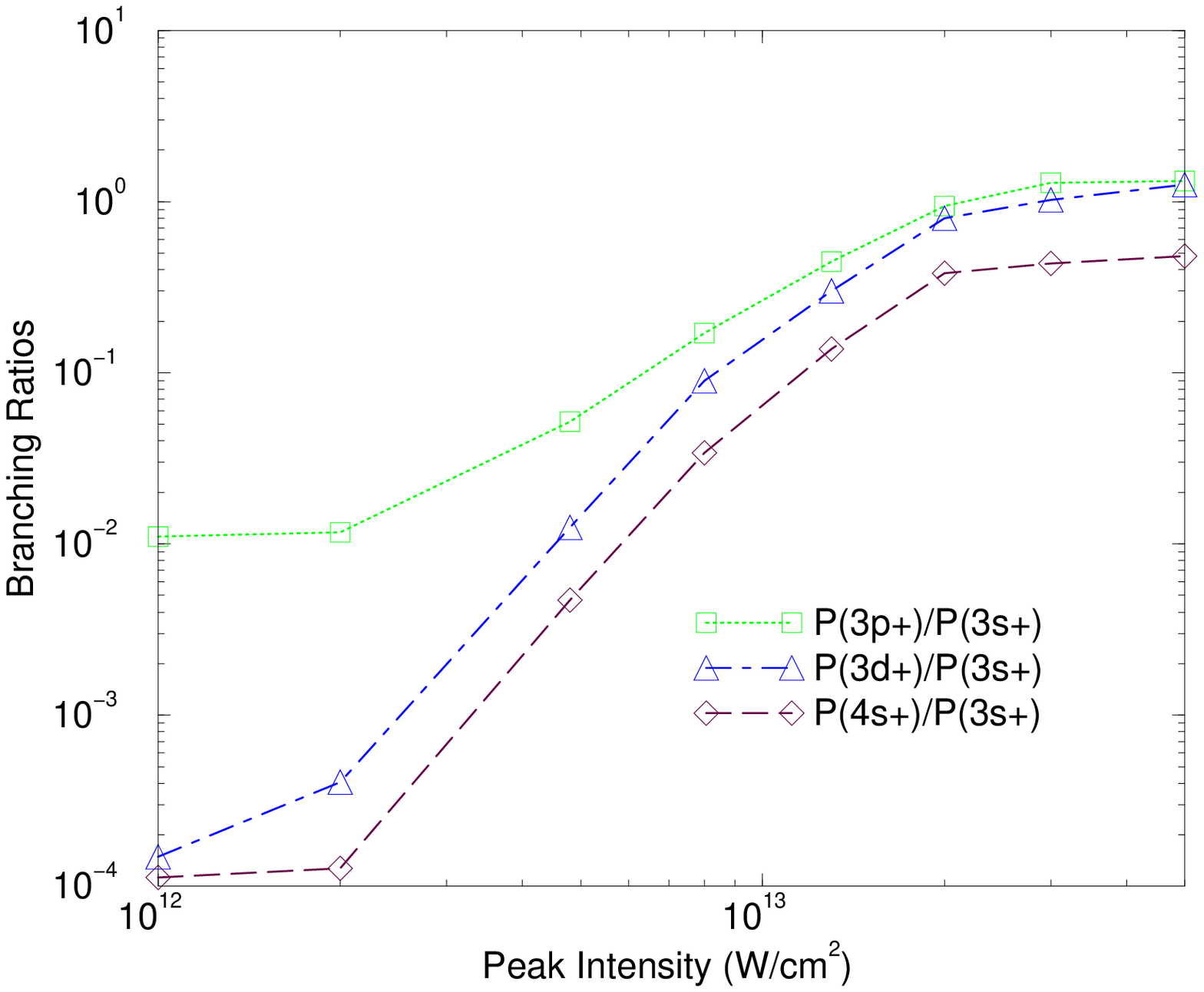}} }
\caption{\label{fig:fig6}}\vspace*{1cm}
\end{figure}

\begin{figure}
\centerline{\resizebox*{8cm}{!}{\includegraphics{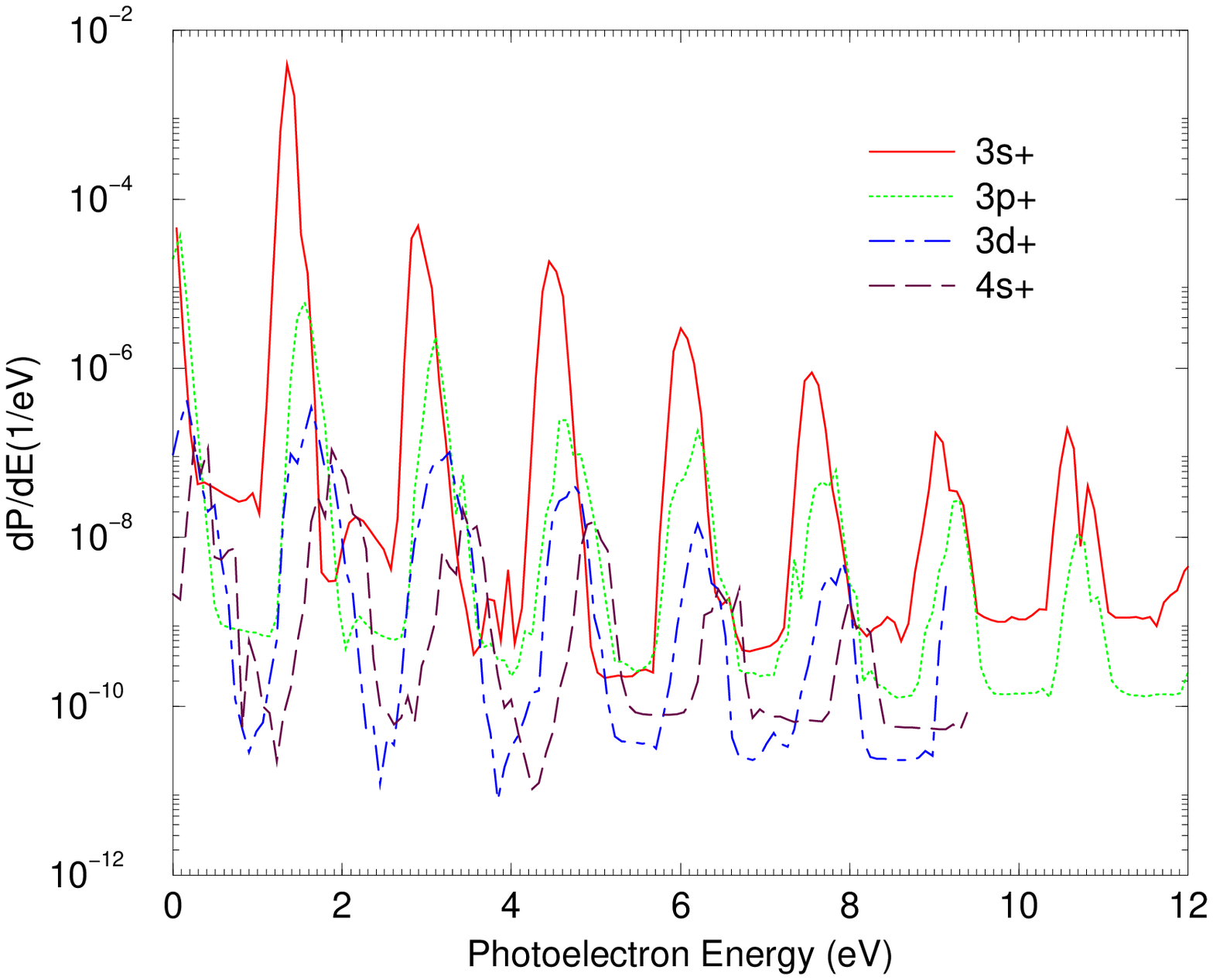}} }
\caption{\label{fig:fig7}}\vspace*{1cm}
\end{figure}

FIG. 7. Branching ratios of Mg$^+$ ions left in different excited
states as a function of laser intensity. The parameters are the same
as in Figure \ref{fig:fig5}.

FIG. 8. Photoelectron ATI energy spectra corresponding to Mg\( ^{+} \)
left in the ground state as well as a few excited states. The laser
peak intensity is \( I_{0}=2\times 10^{12} \) W/cm\( ^{2} \), the
rest of the parameters are the same as in Figure \ref{fig:fig5}.

%%%%%%%%%%%%%%%%%%%%%%%%%%%%%%%%%%%%%%%%%

%%%%%%%%%%%%%%%%%%%%%%%%%%%%%%%%%%%%%%%%%  FIG 1

%%%%%%%%%%%%%%%%%%%%%%%%%%%%%%%%%%%%%%%%%  FIG 2

%%%%%%%%%%%%%%%%%%%%%%%%%%%%%%%%%%%%%%%%%  FIG 3

%%%%%%%%%%%%%%%%%%%%%%%%%%%%%%%%%%%%%%%%%  FIG 3.1

%%%%%%%%%%%%%%%%%%%%%%%%%%%%%%%%%%%%%%%%%  FIG 4

%%%%%%%%%%%%%%%%%%%%%%%%%%%%%%%%%%%%%%%%%  FIG 5
%%%%%%%%%%%%%%%%%%%%%%%%%%%%%%%%%%%%%%%%%  FIG 6

%%%%%%%%%%%%%%%%%%%%%%%%%%%%%%%%%%%%%%%%%  FIG 7

\end{document}